\documentclass[superscriptaddress,twocolumn,showpacs,preprintnumbers,amsmath,amssymb]{revtex4}

\usepackage{graphics}
\usepackage{epsfig}
\usepackage{dcolumn}
\usepackage{bm}

\begin{document}

\title{A note on $\Xi_c(3055)^+$ and $\Xi_c(3123)^+$}

\author{Xiang Liu}\email{xiangliu@pku.edu.cn}
\author{Chong Chen}
\author{Wei-Zhen Deng}
\author{Xiao-Lin Chen}\email{xlchen@th.phy.pku.edu.cn}
\affiliation{Department of Physics, Peking University, Beijing
100871, China}

\date{\today}

\begin{abstract}

Babar Collaboration announced two new excited charmed baryons
$\Xi_c(3055)^+$ and $\Xi_c(3123)^+$. We study their strong decays
assuming they are D-wave states. Some assignments are excluded by
comparing our numerical results with the experimental values of the
total widths of $\Xi_c(3055)^+$ and $\Xi_c(3123)^+$. We also suggest
some possible decay modes, which will be helpful to determine the
properties of $\Xi_c(3055)^+$ and $\Xi_c(3123)^+$.

\end{abstract}


\maketitle

At the recent 2007 Euro-physics Conference on High Energy Physics,
Babar Collaboration reported the preliminary results about the
observations of two new excited charmed baryons $\Xi_c(3055)^+$ and
$\Xi_c(3123)^+$ in the mass distribution of $\Lambda_c^+K^-\pi^+$
\cite{3123-new}. Besides these new observations, Babar also
confirmed the observation of $\Xi_c(2980)^+$ and $\Xi_c(3077)^+$
\cite{babar-2980-3077,belle-2980-3077}. The masses and widths of
$\Xi_c(3055)^+$ and $\Xi_c(3123)^+$ are
\begin{eqnarray*}
&&m_{\Xi_c(3055)^+}=3054.2\pm 1.2\pm0.5\; \mathrm{MeV/c}^2,\\&&
\Gamma_{\Xi_c(3055)^+}=17\pm6\pm11\;\mathrm{MeV/c}^2,\\&&
m_{\Xi_c(3123)^+}=3122.9\pm 1.3\pm0.3\;
\mathrm{MeV/c}^2,\\&&\Gamma_{\Xi_c(3123)^+}=4.4\pm3.4\pm1.7\;\mathrm{MeV/c}^2.
\end{eqnarray*}

In order to understand the recently observed
$\Lambda_c(2880,2940)^+$, $\Xi_c(2980,3077)^{+,0}$, and
$\Omega_c(2768)^0$
\cite{babar-2980-3077,belle-2980-3077,babar-omega,babar-2880,belle-2880},
we studied the strong decays of the S-wave, P-wave, D-wave, and
radially excited charmed baryons using the $^3P_0$ model
systemically \cite{LIU}. (For more details of the $^3P_0$ model, see
Ref.
\cite{Micu,yaouanc,yaouanc-1,yaouanc-book,qpc-1,qpc-2,qpc-90,ackleh,Zou,liu,lujie,baryon-decay}).

{\tiny
\begin{center}
\begin{figure}[htb]
\begin{tabular}{cccc}
\scalebox{0.7}{\includegraphics{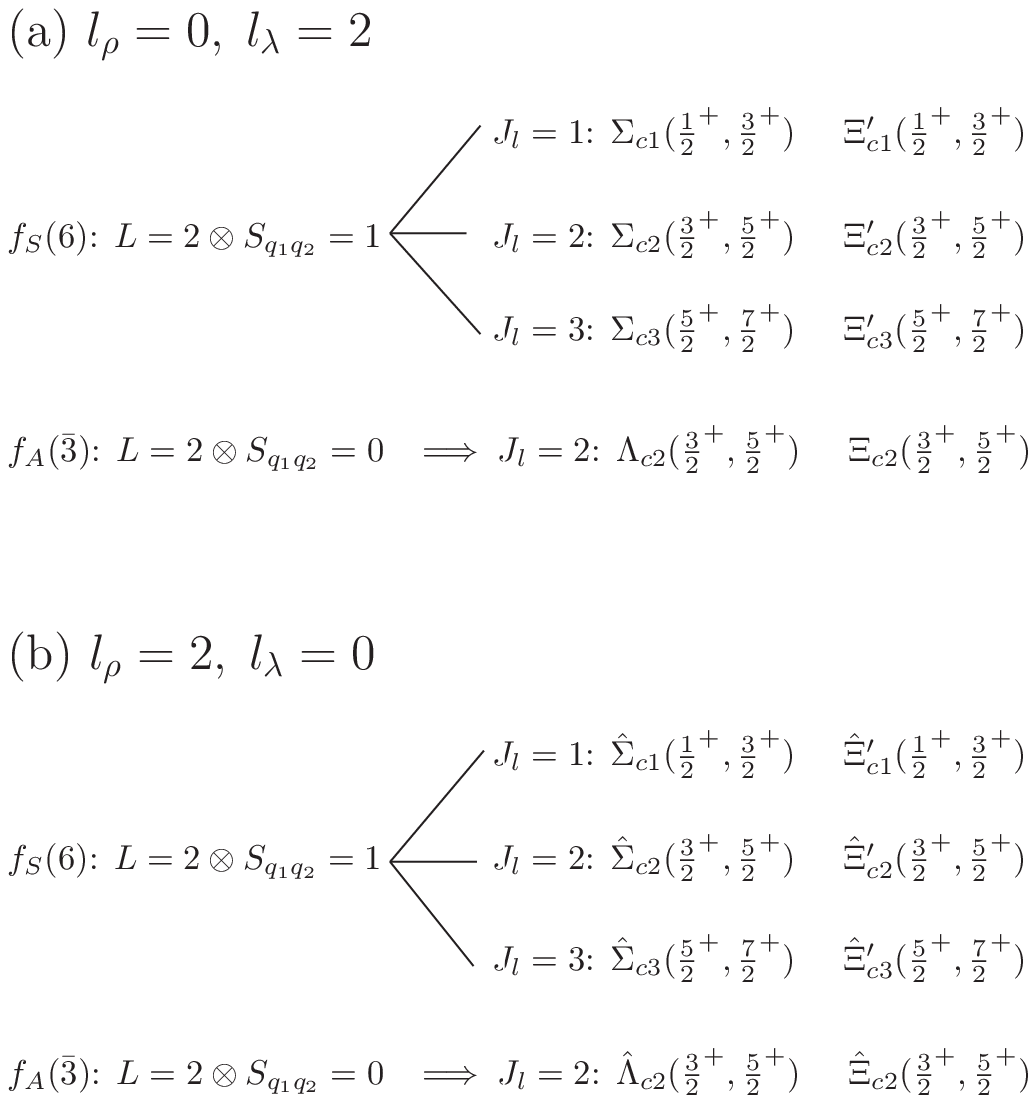}}\\
\scalebox{0.7}{\includegraphics{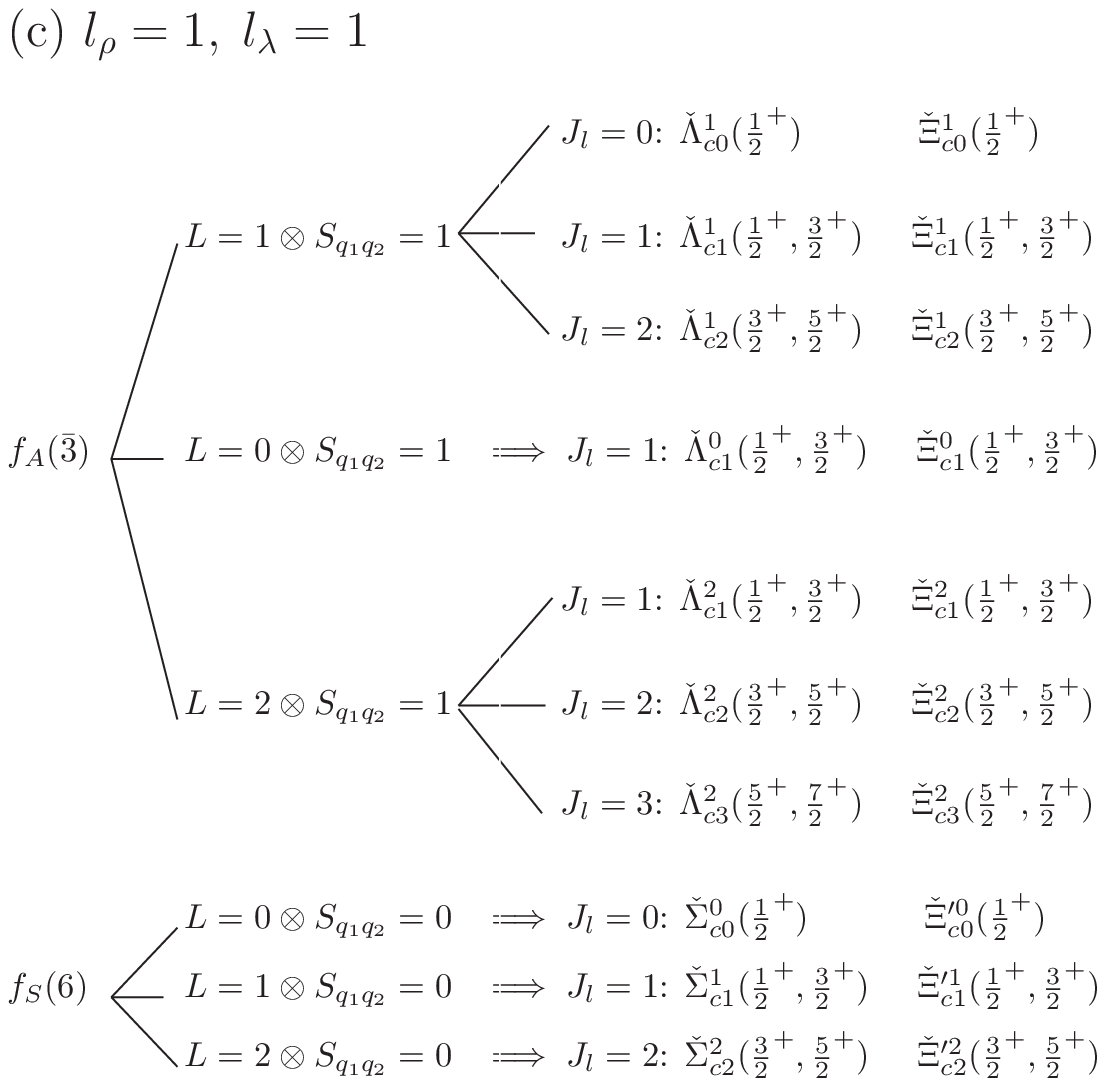}}
\end{tabular}
\caption{The notations for the D-wave charmed
baryons.\label{d-wave}}
\end{figure}
\end{center}
}

In this short note, we analyze the strong decays of $\Xi_c(3055)^+$
and $\Xi_c(3123)^+$ using the same formalism as in Ref. \cite{LIU},
which will be helpful to determine the quantum number of
$\Xi_c(3055)^+$ and $\Xi_c(3123)^+$. Because the parity of these
states is even, they are either the first radial excitation or
D-wave charmed baryons. In our previous work \cite{LIU}, we studied
the total decay width of $\Xi_c(3077)^+$ assuming it's a candidate
of the first radial excitation. Because their masses are close, the
decay pattern of $\Xi_c(3055,3123)^+$ should be similar to that
presented in Ref. \cite{LIU} if either of them is the radial
excitation. In this work, we will not discuss the assignment for
$\Xi_c(3055,3123)^+$ (Interested reader can consult Ref.
\cite{LIU}). In the following, we estimate their strong decays if
$\Xi_c(3055)^+$ and $\Xi_c(3123)^+$ are candidates of D-wave states.
We list the spectrum of D-wave excited spectrum in Fig.
\ref{d-wave}. We omit the detailed expressions of the strong decays
of D-wave charmed baryons derived by this model. Interested readers
may consult our former paper \cite{LIU} for details.

The decay widths of charmed baryons from the $^3P_0$ model involve
several parameters: the strength of quark pair creation from vacuum
$\gamma$, the R value in the harmonic oscillator wave function of
meson and the $\alpha_{\rho,\lambda}$ in the baryon wave functions.
We follow the convention of Ref. \cite{Godfrey} and take $\gamma=
13.4$, which is considered as a universal parameter in the $^3P_0$
model. The R value of $\pi$ and $K$ mesons is $2.1$ GeV$^{-1}$
\cite{Godfrey} while it's $R=2.3$ GeV$^{-1}$ for the $D$ meson
\cite{parameter-2}. $\alpha_{\rho}=\alpha_{\lambda}=0.5$ GeV for the
proton and $\Lambda$ \cite{baryon-decay}. For S-wave charmed
baryons, the parameters $\alpha_{\rho}$ and $\alpha_{\lambda}$ in
the harmonic oscillator wave functions can be fixed to reproduce the
mass splitting through the contact term in the potential model
\cite{potential}. Their values are $\alpha_{\rho}=0.6$ GeV and
$\alpha_{\lambda}=0.6$ GeV. For P-wave and D-wave charmed baryons,
$\alpha_{\rho}$ and $\alpha_{\lambda}$ are expected to lie in the
range $0.5\sim 0.7$ GeV. In the following, our numerical results are
obtained with the typical values
$\alpha_{\rho}=\alpha_{\lambda}=0.6$ GeV. In the following, we
listed the numerical results of the strong decays of $\Xi_c(3055)^+$
and $\Xi_c(3123)^+$ in Table \ref{3055}-\ref{3123}.

At present only total widths of $\Xi_c(3055,3123)^+$ are measured
experimentally. Through comparing our numerical results with
experimental values, we exclude some D-wave assignments for
$\Xi_c(3055,3123)^+$, which are marked by "$\times$" in Table
\ref{3055}-\ref{3123}. In order to fully determine the quantum
numbers of $\Xi_c(3055,3123)^+$, we suggest:

$\bullet$ Search for other possible decay modes of
$\Xi_c(3055,3123)^+$. From Table \ref{3055}-\ref{3123}, one notes
that some decay modes are forbidden for $\Xi_c(3055,3123)^+$ with
several assignments of their quantum numbers, which provides some
useful hint for exclusion or confirmation of certain $J^P$.

$\bullet$ Measure the ratio between different decay modes
$\Xi_c^0\pi^+:\Xi_c'(0)\pi^+:\Xi_c^{*0}\pi^+:
\Sigma_c^{++}K^-:\Sigma_c^{*++}K^-:\Lambda_c^+K^0:D^+\Lambda$. Our
numerical results show this ratio is different for the different
assignment.

\vfill

\section*{Acknowledgments}

This project was supported by the National Natural Science
Foundation of China under Grants 10421503, 10625521, 10705001, and
the China Postdoctoral Science foundation (20060400376). 

\begin{widetext}

\begin{table}[htb]      
\caption{The decay widths of $\Xi^{+}_{c}(3055)$  with different
D-wave assignments. Here we list the results with the typical values
$\alpha_{\rho}=0.6$
 GeV and $\alpha_{\lambda}=0.6$ GeV. \label{3055}}\vskip 0.3cm

\begin{tabular}{l|ccccccccccc}
\hline
 Assignment & \,\,$\Xi^{0}_{c}
\pi^{+}$&\,\,$\Xi'^{0}_{c} \pi^{+}$ & \,\,$\Xi^{\star0}_{c}
\pi^{+}$&\,\,$\Sigma^{++}_{c}k^{-}$&\,\,$\Sigma^{\star++}_{c}k^{-}$&\,\,\,$\Lambda^{+}_{c}\bar{k}^{0}$&$D^{+}\Lambda$&Remark
\\\hline\hline

 $\Xi_{c2}(\frac{3}{2}^{+})$             &$0.0$&  $1.9$  &$0.25$  &$2.2$ &$0.12$ &$0.0$&$0.0$&$$\\
 $\Xi_{c2}(\frac{5}{2}^{+})$             &$0.0$&  $0.028$  &$1.4$  &$0.83\times10^{-2}$ &$0.69 $&$0.0$&$0.0$&$$\\
 $\Xi'_{c1}(\frac{1}{2}^{+})$            &$6.4$&  $1.3$  &$0.38$  &$1.5$ &$0.19$ &$8.0$&$2.4$&$$\\
 $\Xi'_{c1}(\frac{3}{2}^{+})$            &$6.4$&  $0.32$  &$0.96$  &$0.37$ &$0.48$ &$8.0$&$2.4$&$$\\

 $\Xi'_{c2}(\frac{3}{2}^{+})$            &$0.0$&  $2.9$  &$0.36$  &$3.3$ &$0.17$ &$0.0$&$0.0$&$$\\
 $\Xi'_{c2}(\frac{5}{2}^{+})$            &$0.0$&  $0.019$  &$2.1$  &$0.55\times10^{-2}$ &$1.0$ &$0.0$&$0.0$&$$\\

 $\Xi'_{c3}(\frac{5}{2}^{+})$&$0.15$&$0.022$&$0.78\times10^{-2}$&
 $0.63\times10^{-2}$ &$0.30\times10^{-3}$ &$0.18$&$0.0067$&$\times$\\
 $\Xi'_{c3}(\frac{7}{2}^{+})$&$0.15$&$0.012$
 &$0.011$  &$0.35\times10^{-2}$ &$0.41\times10^{-3}$ &$0.18$&$0.0067$&$\times$\\
  \hline
 $\hat{\Xi}_{c2}(\frac{3}{2}^{+})$       &$0.0$&  $27.4$  &$21.3$  &$14.4$ &$2.5$ &$0.0$&$0.0$&$\times$\\
 $\hat{\Xi}_{c2}(\frac{5}{2}^{+})$       &$0.0$&  $27.4$  &$21.3$  &$14.4$ &$2.5$ &$0.0$&$0.0$&$\times$\\
 $\hat{\Xi'}_{c1}(\frac{1}{2}^{+})$      &$163$&  $18.3$  &$3.5$  &$9.6$ &$0.41$ &$205$&$15.5$&$\times$\\
 $\hat{\Xi'}_{c1}(\frac{3}{2}^{+})$      &$163$&  $4.6$  &$8.9$  &$2.4$ &$1.0$ &$205$&$15.5$&$\times$\\

 $\hat{\Xi'}_{c2}(\frac{3}{2}^{+})$      &$0.0$&  $41.1$  &$15.9$  &$21.5$ &$1.9$ &$0.0$&$0.0$&$\times$\\
 $\hat{\Xi'}_{c2}(\frac{5}{2}^{+})$      &$0.0$&  $18.3$  &$24.8$  &$9.6$ &$2.9$ &$0.0$&$0.0$&$\times$\\

 $\hat{\Xi'}_{c3}(\frac{5}{2}^{+})$      &$105$&  $20.9$  &$10.1$  &$10.9$ &$1.2$ &$131$&$10.0$&$\times$\\
 $\hat{\Xi'}_{c3}(\frac{7}{2}^{+})$      &$105$&  $11.7$  &$13.7$  &$6.1$  &$1.6$&$131$&$10.0$&$\times$\\
 \hline
 $\check{\Xi'}^{0}_{c0}(\frac{1}{2}^{+}) $&$0.0$& $0.23$  &$0.46$  &$1.9$ &$2.9$ &$0.0$&$0.0$&$$\\
 $\check{\Xi}^{0}_{c1}(\frac{1}{2}^{+})$ &$9.8$& $0.30$  &$0.15$  &$2.6$&$0.95$  &$12.4$&$0.60$&$$\\
 $\check{\Xi}^{0}_{c1}(\frac{3}{2}^{+})$ &$9.8$& $0.075$  &$0.38$  &$0.65$&$2.4$  &$12.4$&$0.60$&$$\\
 $\check{\Xi'}^{1}_{c1}(\frac{1}{2}^{+})$&$0.0$&  $34.7$  &$9.8$  &$36.0$ &$4.1$ &$0.0$&$0.0$&$\times$\\
 $\check{\Xi'}^{1}_{c1}(\frac{3}{2}^{+})$&$0.0$&  $8.7$  &$24.4$  &$9.0$ &$10.3$ &$0.0$&$0.0$&$\times$\\
 $\check{\Xi}^{1}_{c0}(\frac{1}{2}^{+})$ &$0.0$&  $34.7$  &$39.1$  &$36.0$ &$16.6$ &$0.0$&$0.0$&$\times$\\
 $\check{\Xi}^{1}_{c1}(\frac{1}{2}^{+})$ &$97.6$& $17.4$  &$4.9$  &$18.0$ &$2.1$ &$122$&$28.2$&$\times$\\
 $\check{\Xi}^{1}_{c1}(\frac{3}{2}^{+})$ &$97.6$& $4.3$  &$12.2$  &$4.5$ &$5.2$ &$122$&$28.2$&$\times$\\
 $\check{\Xi}^{1}_{c2}(\frac{3}{2}^{+})$ &$0.0$&  $21.7$  &$2.4$  &$22.5$ &$1.0$ &$0.0$&$0.0$&$\times$\\
 $\check{\Xi}^{1}_{c2}(\frac{5}{2}^{+})$ &$0.0$&$0.0$  &$14.7$  &$0.0$ &$6.2$ &$0.0$&$0.0$&$$\\
 $\check{\Xi'}^{2}_{c2}(\frac{3}{2}^{+})$&$0.0$&  $8.6$  &$4.7$  &$12.3$&$1.5$  &$0.0$&$0.0$&$$\\
 $\check{\Xi'}^{2}_{c2}(\frac{5}{2}^{+})$&$0.0$&  $4.7$  &$8.7$  &$2.8$ &$4.4$ &$0.0$&$0.0$&$$\\
 $\check{\Xi}^{2}_{c1}(\frac{1}{2}^{+})$ &$21.9$&  $5.7$  &$2.0$  &$8.2$ &$1.1$ &$27.2$&$12.2$&$\times$\\
 $\check{\Xi}^{2}_{c1}(\frac{3}{2}^{+})$ &$21.9$&  $1.4$  &$4.9$  &$2.1$ &$2.8$ &$27.2$&$12.2$&$\times$\\
 $\check{\Xi}^{2}_{c2}(\frac{3}{2}^{+})$ &$0.0$&  $12.9$  &$4.1$  &$18.5$ &$1.5$ &$0.0$&$0.0$&$\times$\\
 $\check{\Xi}^{2}_{c2}(\frac{5}{2}^{+})$ &$0.0$&  $3.2$  &$11.7$  &$1.9$ &$6.3$ &$0.0$&$0.0$&$$\\
 $\check{\Xi}^{2}_{c3}(\frac{5}{2}^{+})$ &$17.4$&  $3.6$  &$1.9$  &$2.2$  &$0.41$&$21.9$&$2.1$&$\times$\\
 $\check{\Xi}^{2}_{c3}(\frac{7}{2}^{+})$ &$17.4$&  $2.0$  &$2.5$  &$1.2$  &$0.57$&$21.9$&$2.1$&$\times$\\
 \hline\hline
\end{tabular}
\end{table}

\begin{table}[htb]      
\caption{The decay widths of $\Xi^{+}_{c}(3123)$  with different
D-wave assignments. Here we list the results with the typical values
$\alpha_{\rho}=0.6$
 GeV and $\alpha_{\lambda}=0.6$ GeV. \label{3123}}\vskip 0.3cm
\begin{tabular}{l|ccccccccccc}
\hline
 Assignment & \,\,$\Xi^{0}_{c}
\pi^{+}$&\,\,$\Xi'^{0}_{c} \pi^{+}$ & \,\,$\Xi^{\star0}_{c}
\pi^{+}$&\,\,$\Sigma^{++}_{c}k^{-}$&\,\,$\Sigma^{\star++}_{c}k^{-}$&\,\,\,$\Lambda^{+}_{c}\bar{k}^{0}$&$D^{+}\Lambda$&Remark
\\\hline\hline

 $\Xi_{c2}(\frac{3}{2}^{+})$             &$0.0$&  $2.8$  &$0.43$  &$4.5$&$0.49$  &$0.0$&$0.0$&$$\\
 $\Xi_{c2}(\frac{5}{2}^{+})$             &$0.0$&  $0.075$  &$2.3$  &$0.053$ &$2.8$&$0.0$&$0.0$&$$\\
 $\Xi'_{c1}(\frac{1}{2}^{+})$            &$8.3$&  $1.9$  &$0.63$  &$3.0$ &$0.79$ &$10.2$&$5.5$&$\times$\\
 $\Xi'_{c1}(\frac{3}{2}^{+})$            &$8.3$&  $0.46$  &$1.6$  &$0.76$&$2.0$  &$10.2$&$5.5$&$\times$\\

 $\Xi'_{c2}(\frac{3}{2}^{+})$            &$0.0$&  $4.2$  &$0.60$  &$6.8$&$0.72$  &$0.0$&$0.0$&$$\\
 $\Xi'_{c2}(\frac{5}{2}^{+})$            &$0.0$&  $0.050$  &$3.4$  &$0.035$&$4.3$  &$0.0$&$0.0$&$$\\

 $\Xi'_{c3}(\frac{5}{2}^{+})$&$0.32$&$0.057$&$0.026$&
 $0.040$ &$0.010$ &$0.44$&$0.069$&$\times$\\
 $\Xi'_{c3}(\frac{7}{2}^{+})$&$0.32$&  $0.032$
 &$0.035$  &$0.023$&$0.013$  &$0.44$&$0.069$&$\times$\\
  \hline
 $\hat{\Xi}_{c2}(\frac{3}{2}^{+})$       &$0.0$&  $58.9$  &$53.0$  &$56.0$&$30.0$  &$0.0$&$0.0$&$\times$\\
 $\hat{\Xi}_{c2}(\frac{5}{2}^{+})$       &$0.0$&  $58.9$  &$53.0$  &$56.0$&$30.0$  &$0.0$&$0.0$&$\times$\\
 $\hat{\Xi'}_{c1}(\frac{1}{2}^{+})$      &$311$&  $39.2$  &$8.8$  &$37.4$ &$5.0$ &$411$&$85.9$&$\times$\\
 $\hat{\Xi'}_{c1}(\frac{3}{2}^{+})$      &$311$&  $9.8$  &$22.1$  &$9.3$&$12.5$  &$411$&$85.9$&$\times$\\

 $\hat{\Xi'}_{c2}(\frac{3}{2}^{+})$      &$0.0$&  $88.3$  &$40.0$  &$84.1$&$22.5$  &$0.0$&$0.0$&$\times$\\
 $\hat{\Xi'}_{c2}(\frac{5}{2}^{+})$      &$0.0$&  $39.2$  &$61.8$  &$37.4$&$35.0$  &$0.0$&$0.0$&$\times$\\

 $\hat{\Xi'}_{c3}(\frac{5}{2}^{+})$      &$200$&  $44.8$  &$25.2$  &$42.7$&$14.3$  &$264$&$55.3$&$\times$\\
 $\hat{\Xi'}_{c3}(\frac{7}{2}^{+})$      &$200$&  $25.2$  &$34.0$  &$24.0$&$19.3$  &$264$&$55.3$&$\times$\\
 \hline
 $\check{\Xi'}^{0}_{c0}(\frac{1}{2}^{+}) $&$0.0$& $4.3$  &$0.35$  &$0.015$&$2.5$  &$0.0$&$0.0$&$$\\
 $\check{\Xi}^{0}_{c1}(\frac{1}{2}^{+})$ &$36.5$& $5.8$  &$0.12$  &$0.020$&$0.82$ &$52.7$&$1.8$&$\times$\\
 $\check{\Xi}^{0}_{c1}(\frac{3}{2}^{+})$ &$36.5$& $1.4$  &$0.29$  &$0.005$&$2.0$ &$52.7$&$1.8$&$\times$\\
 $\check{\Xi'}^{1}_{c1}(\frac{1}{2}^{+})$&$0.0$&  $54.3$  &$17.0$  &$80.2$&$18.3$  &$0.0$&$0.0$&$\times$\\
 $\check{\Xi'}^{1}_{c1}(\frac{3}{2}^{+})$&$0.0$&  $13.6$  &$42.6$  &$20.1$&$45.8$  &$0.0$&$0.0$&$\times$\\
 $\check{\Xi}^{1}_{c0}(\frac{1}{2}^{+})$ &$0.0$&  $54.3$  &$68.2$  &$80.2$&$73.3$  &$0.0$&$0.0$&$\times$\\
 $\check{\Xi}^{1}_{c1}(\frac{1}{2}^{+})$ &$139$& $27.1$  &$8.5$  &$40.1$&$9.2$  &$175$&$73.3$&$\times$\\
 $\check{\Xi}^{1}_{c1}(\frac{3}{2}^{+})$ &$139$& $6.8$  &$21.3$  &$10.0$ &$22.9$ &$175$&$73.3$&$\times$\\
 $\check{\Xi}^{1}_{c2}(\frac{3}{2}^{+})$ &$0.0$&  $33.9$  &$4.3$  &$50.1$&$4.6$  &$0.0$&$0.0$&$\times$\\
 $\check{\Xi}^{1}_{c2}(\frac{5}{2}^{+})$ &$0.0$&$0.0$  &$25.6$  &$0.0$&$27.5$  &$0.0$&$0.0$&$\times$\\
 $\check{\Xi'}^{2}_{c2}(\frac{3}{2}^{+})$&$0.0$&  $10.5$  &$10.0$  &$21.3$&$8.1$  &$0.0$&$0.0$&$\times$\\
 $\check{\Xi'}^{2}_{c2}(\frac{5}{2}^{+})$&$0.0$&  $9.9$  &$13.9$  &$9.8$&$16.9$  &$0.0$&$0.0$&$\times$\\
 $\check{\Xi}^{2}_{c1}(\frac{1}{2}^{+})$ &$21.6$&  $7.0$  &$2.8$  &$14.2$&$4.0$  &$25.2$&$21.1$&$\times$\\
 $\check{\Xi}^{2}_{c1}(\frac{3}{2}^{+})$ &$21.6$&  $1.7$  &$7.1$  &$3.6$&$10.0$  &$25.2$&$21.1$&$\times$\\
 $\check{\Xi}^{2}_{c2}(\frac{3}{2}^{+})$ &$0.0$&  $15.7$  &$8.1$  &$32.0$&$7.4$  &$0.0$&$0.0$&$\times$\\
 $\check{\Xi}^{2}_{c2}(\frac{5}{2}^{+})$ &$0.0$&  $6.6$  &$17.7$  &$6.6$&$23.2$  &$0.0$&$0.0$&$\times$\\
 $\check{\Xi}^{2}_{c3}(\frac{5}{2}^{+})$ &$32.7$&  $7.5$  &$4.4$  &$7.5$&$3.0$  &$43.1$&$10.1$&$\times$\\
 $\check{\Xi}^{2}_{c3}(\frac{7}{2}^{+})$ &$32.7$&  $4.2$  &$5.9$  &$4.2$&$4.1$  &$43.1$&$10.1$&$\times$\\
 \hline\hline
\end{tabular}
\end{table}
\end{widetext}

\end{document}